\documentclass[aps,prc,showpacs,twocolumn,amsmath,amssymb,superscriptaddress]{revtex4}
\usepackage{graphicx}

\begin{document}

\title{New Measurements and Quantitative Analysis of Electron
  Backscattering in the Energy Range of Neutron $\beta$-Decay}

\affiliation{W. K. Kellogg Radiation Laboratory, California Institute
of Technology, Pasadena, CA 91125}
\affiliation{North Carolina State University, Raleigh, NC 27695}
\affiliation{CENPA, University of Washington, Seattle, WA 98195}
\affiliation{Physics Department, University of Winnipeg, Winnipeg, MB
  R3B 2E9 CANADA}

\author{J.W.~Martin}
\affiliation{W. K. Kellogg Radiation Laboratory, California Institute
of Technology, Pasadena, CA 91125}
\affiliation{Physics Department, University of Winnipeg, Winnipeg, MB
  R3B 2E9 CANADA}

\author{J.~Yuan}
\affiliation{W. K. Kellogg Radiation Laboratory, California Institute
of Technology, Pasadena, CA 91125}

\author{M.J.~Betancourt}
\affiliation{W. K. Kellogg Radiation Laboratory, California Institute
of Technology, Pasadena, CA 91125}

\author{B.W.~Filippone}
\affiliation{W. K. Kellogg Radiation Laboratory, California Institute
of Technology, Pasadena, CA 91125}

\author{S.A.~Hoedl}
\affiliation{CENPA, University of Washington, Seattle, WA 98195}

\author{T.M.~Ito}
\altaffiliation[Present address: ]{Los Alamos National Laboratory, Los
Alamos, NM 87545}
\affiliation{W. K. Kellogg Radiation Laboratory, California Institute
of Technology, Pasadena, CA 91125}

\author{B.~Plaster}
\affiliation{W. K. Kellogg Radiation Laboratory, California Institute
of Technology, Pasadena, CA 91125}

\author{A.R.~Young}
\affiliation{North Carolina State University, Raleigh, NC 27695}

\begin{abstract}
We report on the first detailed measurements of electron
backscattering from plastic scintillator targets, extending our
previous work on beryllium and silicon targets.  The scintillator
experiment posed several additional experimental challenges associated
with charging of the scintillator target, and those challenges are
addressed in detail.  In addition, we quantitatively compare the
energy and angular distributions of this data, and our previous data,
with electron transport simulations based on the Geant4 and Penelope
Monte Carlo simulation codes.  The Penelope simulation is found
globally to give a superior description of the data.  Such information
is crucial for a broad array of weak-interaction physics experiments,
where electron backscattering can give rise to the dominant
detector-related systematic uncertainty.
\end{abstract}


\maketitle

\section{Introduction}

Our previous work on backscattering \cite{bib:prcme} from beryllium
and silicon extended the work of others
\cite{bib:gerard,bib:massoumi1,bib:massoumi2} to energies relevant for
nuclear physics applications.  We further extend this work to a more
relevant target, organic scintillator.  The main experimental
challenge arose from the fact that the target material is not
conducting, necessitating mitigation of effects due to charging in
order to accurately sense currents.

\section{Experiment Overview}

As for our previous measurements, the experiment consisted of an
electron gun and a scattering chamber containing a movable target
\cite{bib:prcme}.  Two modes of acquiring backscattering data were
used.  In one mode a silicon detector was used to detect the energy
and angle of backscattered electrons.  In a second higher-current
mode, the electrical current due to the backscattered electrons
incident on the chamber walls was measured.  These two modes were
referred to as silicon detector mode and current integration mode,
respectively.

The plastic scintillator targets used for the experiment were obtained
from Eljen Technologies \cite{bib:eljen}.  The type of plastic
scintillator used was EJ-204.  The density of EJ-204 is 1.032
g/cm$^3$, with the dominant elemental composition being $5.21\times
10^{22}$ H atoms/cm$^3$, and $4.74\times 10^{22}$ C atoms/cm$^3$.  A
sample of plastic scintillator was coated with a 500 \AA~thick layer
of aluminum via evaporation.  Targets of a suitable size and shape
were then cut from this sample, with care taken to preserve the
aluminized front face.  The resulting targets were 1''$\times$1''
square, 3~mm thick, with a thinner tab of aluminized plastic
projecting from the top for suspension from a central rotatable
feedthrough.

As before, the energy and angular distribution of the backscattered
electrons was measured using a 25~mm$^2$ active area silicon detector.
The energy resolution of the silicon detector system was typically
$\sigma=4.3$~keV, and was independent of energy.

\section{Backscattering Measurements}

Backscattering measurements were performed for normal incidence upon
the target.  Measurements were performed for incident electron
energies of 43.5, 63.9, 83.8, 104, and 124 keV.  For each energy, both
silicon detector mode and current integration mode measurements were
taken.

\subsection{Silicon Detector Mode}

The quantities which varied for the silicon detector measurements
conducted at a particular electron beam energy $E_{\rm beam}$ were the
detection angle $\theta$, and the energy $E$ the backscattered
electron deposited in the silicon detector.  The angle $\theta$ was
defined relative to the surface normal vector of the material, so that
$\theta=0$~degrees corresponds to a backscattering event where the
electron went directly back along the incident beam direction.  The
dimensionless energy $q=E/E_{\rm beam}$ will also be used.

For silicon detector mode, the systematic uncertainty is unchanged
from our previous work, resulting in an average 12\% normalization
systematic for each angular setting of the detector.  However, we
specifically addressed current detection and charging of the target,
as will now be discussed.

To ensure the reliability of measurements of current, we compared the
measurements of two different, carefully calibrated current
integrators against one another for identical experimental conditions,
and in turn cross-compared those measurements against two calibrated
picoammeters.  The measurements were consistent at the 3\% level.

The possible effects of charging and incomplete current detection were
monitored by observing scintillation light from the electron beam as
it struck the scintillator target, on a camera mounted behind the
target outside a viewport in the chamber (at $\theta=180^\circ$).
Over the course of taking a complete angular range of data for a
particular incident beam energy, the brightness of the scintillation
light did not visibly change with time.  Previous tests with uncoated
scintillator, or scintillator coated poorly with graphite, had shown
that the emanation of scintillation light from the target would
eventually cease, indicating that the electron beam had been diverted
away from the central spot on the target.  Also discharges would be
seen due to arcing from the face of the scintillator to the conducting
target rod.  No such effects could be seen with the Al-coated
scintillator target.  On leaving the beam on the target for long
periods of time, no change in the sensed current was seen at the 1\%
level.  To also search for charging, the electron beam could be
switched off and on rapidly by inserting a Faraday cup upstream of the
chamber.  Upon restoration of the beam, the current sensed by the
target was found to agree with the value before intercepting the beam
with the Faraday cup to the 1\% level.

The effect of the aluminum layer can be estimated by comparing the
thickness of the Al layer to the mean range of the electrons.  The
mean range for 43.5~keV electrons in plastic scintillator is 30~$\mu$m
\cite{bib:estar}, which is three orders of magnitude larger than the
thickness of the Al layer.  This implies that scattering from the Al
is suppressed below the 1\% level, even for the lowest energy incident
electrons reported in this work.  Both estimates were confirmed in
Monte Carlo studies of scattering from thin layers \cite{bib:seth}.
For the silicon detector data, the effects of the Al layer were
therefore neglected.

As a cross-check of the reproducibility, dead-time, and current
detection uncertainties due to charging, runs were taken for various
beam currents.  The normalized yield for these runs was found not to
vary outside the previously quoted 7\% reproducibility uncertainty.

For completeness, a catalog of the dominant systematic uncertainties
considered for silicon detector mode is displayed in
Table~\ref{tab:simode}.
\begin{table}[ht]
\begin{ruledtabular}
\begin{tabular}{lc}
Effect               & Uncertainty \\
\hline
Reproducibility      & 7\% \\
Active Area          & 4\% \\
Finite Beam Spot     & 5\%$\times\sin\theta$ \\
Dead Time            & 3\% \\
Alignment            & 2\% \\
Current Detection    & 3\% \\
Target Deterioration & 1\% \\
\hline
Extrapolation over $q$ & 6\%-20\% \\
Extrapolation over $\theta$ & 4\% \\
\hline
Total                & 12-23\%
\end{tabular}
\end{ruledtabular}
\caption{Dominant systematic uncertainties for silicon detector mode.
\label{tab:simode}}
\end{table}
Aside from target deterioration, which will be discussed in the next
section, these systematic uncertainties were described in detail in
our previous work~\cite{bib:prcme}.  The effects listed in the upper
portion of Table~\ref{tab:simode} refer to results for the observable
$\frac{1}{N_e}\frac{dN}{d\Omega dq}$, and average 12\%.
The effects in the lower portion of
Table~\ref{tab:simode} refer to extrapolation uncertainties
encountered when integrating these data over $q$ and/or $\theta$, and
these must be added in quadrature to those above when considering our
results for $d\eta/d\Omega$ and $\eta$ from silicon detector mode.
For extrapolation over $q$, the uncertainty is larger for lower beam
energy, as the finite detection threshold becomes more important;
hence the 20\% value refers to beryllium at 43.5~keV incident energy,
since the beryllium data is more peaked at lower $q$, while the 6\%
value refers to data taken with 124~keV incident energy.  Considering
the range given for the total systematic uncertainty, the lower bound
(12\%) refers to all observations of $\frac{1}{N_e}\frac{dN}{d\Omega
dq}$, while the upper bound (23\%) refers specifically to the
extraction of $\eta$ from beryllium at 43.5~keV, where extrapolation
uncertainties dominate.

For these experiments, the silicon detector threshold varied from
10~keV to 18~keV depending on noise conditions at the time of any
given data-taking run.

Effects due to backgrounds from X-rays, and multiply backscattered
electrons were also discussed in detail in our previous work.
Multiply backscattered electrons resulted in systematic uncertainties
at the 3-5\% level at $q=0.2$ for the highest beam energy considered.
But this contribution to the systematic uncertainty is below 1\% above
$q=0.3$, and is negligible for lower beam energies.

\subsection{Current Integration Mode}

For each beam energy setting, current integration mode measurements
were also performed.

The largest potential systematic effect, which was previously
uncontrolled, arose from deterioration of the scintillator target by
the electron beam.  Over the course of an hour, running at beam
currents of tens of nA, $\eta$ was observed to steadily increase with
time, plateauing at a value typically 15\% larger than its initial
value.  The transition to larger $\eta$ was found to occur more
rapidly with higher beam currents and higher beam energies.  The glow
of the beam spot on the target was also monitored on the video camera
and found to decrease in brightness in a correlated way.  The
brightness was found not to recover after leaving the scintillator in
vacuum over several days, ruling out charging/discharging of the bulk
scintillator.  Upon removal of an affected target from the vacuum
chamber, a small brownish spot within the scintillator could be
observed, with no obvious deterioration of the mirror-like aluminized
front face.  We believe the discoloration, reduction in scintillation
light, and increase in $\eta$ are symptoms of a chemical change in the
scintillator, possibly resulting in the liberation of hydrogen.  This
would increase the relative carbon content and hence $\eta$.

To control this potential systematic, new scintillator targets were
used and were exposed to electron beams with currents less than 1~nA.
The targets were exposed to beam for the minimum time possible for
currents to be sensed accurately using picoammeters.  This limited the
contribution to the systematic uncertainty to less than 1\%, based on
the slope measured during the prior hour-long measurements.

A dependence on beam current at the 3\% level had been seen for our
previous current integration mode data on Be and Si
target~\cite{bib:prcme}.  For those data, the current dependence was
attributed to charging of various components in the chamber.  As our
new studies were done generally at lower beam currents, we expect this
contribution to be smaller, but retain the pessimistic upper limit of
3\%.

In order to characterize and control the effects of low-energy
secondary electrons (for our work, defined to be below 50 eV in
energy), a wire mesh cage was inserted into the chamber and held at
-55~V.  From an electrostatic model of the electric potential in the
chamber, this resulted in a minimum potential wall of -50~V between
the target and the chamber.  For the purposes of modelling these
measurements, the integrated backscatter current was therefore taken
to be due to all electrons emanating from the target with greater than
50 eV.

The wire mesh cage was referred to as ``the grid'' consistent with
terminology used in previous backscattering literature.  The grid was
constructed from a thin copper rod bent into a cylindrical shape.
Steel wire was wound on the resultant frame in an end-over-end pattern
resulting in vertical wires equally spaced running down the sides of
the cylindrical shape.  The resultant cylindrical wire cage resembled
a bird cage which enclosed the target.  In this way, biasing the grid
at negative potential prevented secondaries from traveling from the
chamber walls to the target and vice versa.

However, as with our previous work, the insertion of the grid resulted
in systematic effects due to incomplete compensation for secondaries
and due to effects of backscattered electrons striking the grid
itself.

A residual correction due to a piece of conducting target rod
penetrating the top of the grid must be applied.  The rod subtended a
small but finite solid angle viewing the beam spot on the target.
High-energy backscattered electrons could strike that portion of the
rod, resulting in secondary electron production.  Secondary electrons
created on that portion of the rod would not be suppressed by the grid
and would be collected on the target.  This resulted in a residual
dependence of the apparent $\eta$ on target voltage, which could be
corrected.  The correction gave rise to a 7\% contribution to the
systematic uncertainty in the previous work, and was the dominant
uncertainty.  For this work, the contribution was reduced to 3\% by
reducing the solid angle subtended by the relevant portion of the
target rod.  This reduced the size of the correction, and hence the
systematic uncertainty.  Detailed analytical calculations of the
effect of the solid angle and its variation with e.g. distance of the
beam spot from the top of the target close to the target rod gave
additional confidence in this uncertainty.  As before, secondaries
created on the grid were characterized by monitoring the grid current,
and contributed at the 1\% level.

A potential systematic effect arises due to differences in backscatter
yields above 50 eV because of the presence of the Al coating on the
target.  Electrons with energies above 100 eV could not be adequately
assessed via altering the voltage on the grid, while from the electron
transport arguments presented earlier, only for energies above about
10 keV can the effects of the Al layer be argued to be negligible.
From similar considerations to the extrapolation uncertainty for
silicon detector mode, we limit possible effects of the Al coating in
this energy range to the 6\% level.

The dominant systematic uncertainties for current-integration mode
results for our new scintillator target data are listed in
Table~\ref{tab:currmode}.
\begin{table}[ht]
\begin{ruledtabular}
\begin{tabular}{lc}
Effect               & Uncertainty \\
\hline
Target rod correction& 3\% \\
Grid secondaries     & 1\% \\
Reproducibility      & 5\% \\
Current dependence   & 3\% \\
Target deterioration & 1\% \\
Al coating effects   & 6\% \\
\hline
Total                & 9\%
\end{tabular}
\end{ruledtabular}
\caption{Dominant systematic uncertainties for current integration mode for
scintillator target data.\label{tab:currmode}}
\end{table}

\section{Results}

\subsection{Silicon Detector Mode}

\begin{figure}
\includegraphics[width=0.5\textwidth]{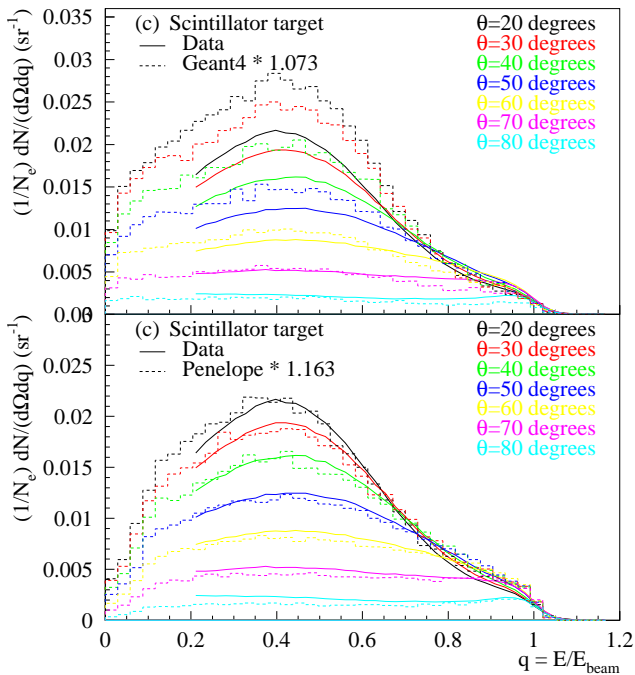}
\caption{(Color online) Normal incidence backscattering from
scintillator target at $E_{\rm beam}=124$~keV.  Curves represent data
taken with silicon detector.  Histogram is Monte Carlo simulation
based on (a) Geant4, and (b) Penelope.  Systematic uncertainty in the
normalization of the data is estimated to be 12\% on average, ranging
from 11\% at small angles to 15\% at large angles.  A scale factor of
1.073 is applied to the (a) Geant4 simulation.  In (b), the factor is
1.163 for Penelope.\label{fig:120}}
\end{figure}

The normalized, background-subtracted spectra accumulated for various
detector angles for 124~keV electrons normally incident on a
scintillator target is shown in Fig.~\ref{fig:120}.
The data are plotted as a function of the dimensionless energy $q$.
On the vertical axis, $\frac{1}{N_e}\frac{dN}{dqd\Omega}$, the number
of counts per incident electron, per unit $q$, per unit solid angle is
plotted.  In the absence of the effects of detector response
(resolution and backscattering), this would be the normal-incidence
backscattered fraction per unit $q$, per unit solid angle.

Monte Carlo simulations of backscattering were carried out with custom
codes based on the Geant4 \cite{bib:geant4nim} and Penelope
\cite{bib:penelope2} toolkits.  The custom aspects of the simulation
codes were described in Ref.~\cite{bib:prcme}, for example, their
handling of backscattering from the silicon detector itself and the
detector response function (relating to this particular observable).
Fig.~\ref{fig:120} compares the data with the results of these two
codes.  In each case, a scale factor is applied to the Monte Carlo.
The scale factor was determined from a fit to the data, which is
described in Section~\ref{sec:fit}.  As for our previous work, Geant4
systematically underestimates the peak in the data near $q=0.95$.
However the positions in $q$ of the low-energy and elastic peaks are
rather well-described by Geant4.  In the case of the Penelope
simulation, when the Monte Carlo is rescaled, it is apparent that
trends in both energy and angle are well represented by Penelope.
This result is discussed quantitatively in Section~\ref{sec:fit}.

\begin{figure}
\includegraphics[width=0.5\textwidth]{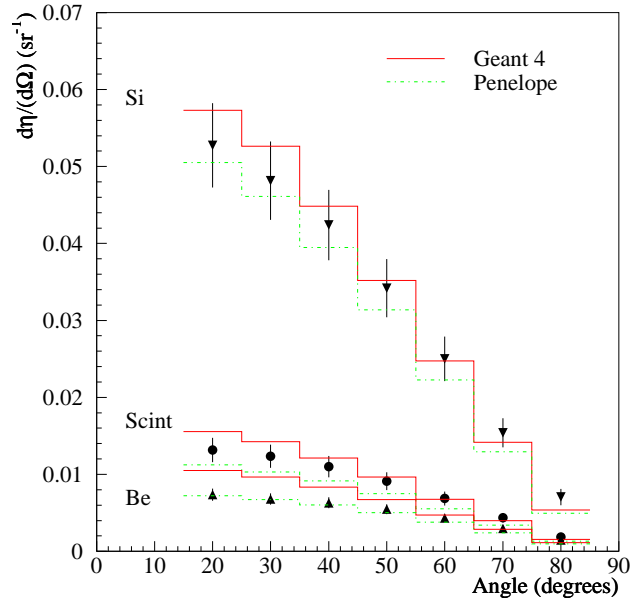}
\caption{(Color online) d$\eta$/d$\Omega$ for beryllium (triangles)
and silicon (inverted triangles) targets at $E_{\rm beam}=124$~ keV.
Black points with error bars indicate data with total normalization
systematic uncertainties shown.  Red solid histogram indicates the
results of the Geant4-based Monte Carlo simulation.  Green dot-dashed
histogram indicates the results of the Penelope-based Monte Carlo
simulation.  No Monte Carlo scale factors are
included.\label{fig:detado}}
\end{figure}

The angular dependence of the backscattered fraction can be determined
by integrating the data over $q$.  The result of doing so is shown in
Fig.~\ref{fig:detado}, and is compared with our previous results for
silicon and beryllium targets.
A linear fit based on the first 20 keV of data above the analysis cut
was used to extrapolate to 50 eV (the defined threshold for secondary
electrons), so that these integrals and subsequent integrals could be
compared with the current integration measurements.  An additional
systematic uncertainty was assigned to the extrapolation, based on
differences of fit functions, and comparison to models.  For 124 keV
beam energy, this extra systematic uncertainty was 5\%, resulting in a
contribution due to extrapolation of typically 6\% (see
Table~\ref{tab:simode}).

Fig.~\ref{fig:detado} also compares the Geant4 and Penelope
simulations with the data.  No Monte Carlo scale factors are applied
for this comparison.  Each Monte Carlo separately correctly predicts
that the scintillator results should be larger than the beryllium
results, as expected due to the larger $Z$ of the carbon nuclei in
scintillator.  As noted previously, the Penelope simulation tends to
better describe trends in angle.  The Geant4 distributions are
somewhat narrower compared to the data and Penelope.  Additionally,
the Geant4 simulation gives systematically larger backscattering from
each material than does the Penelope simulation.

\begin{figure}
\includegraphics[width=0.5\textwidth]{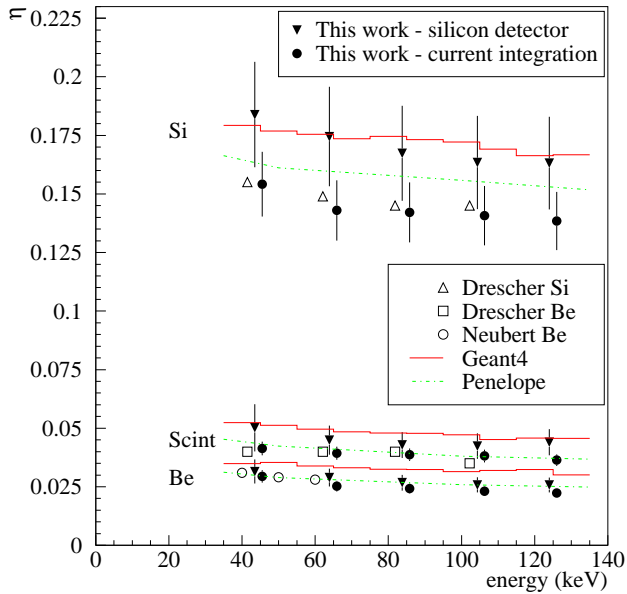}
\caption{(Color online) Normal incidence backscattering from Be, Si,
and plastic scintillator targets.  Integrated silicon detector
measurements are shown by the inverted filled triangles.  Current
integration measurements are shown by filled circles.  Total
systematic uncertainties are shown and the current integration
measurements are displaced by 2 keV so that the error bars do not
overlap.  Previous current integration
measurements~\cite{bib:Drescher,bib:Neubert1} are displayed.  The
histograms show the results of the Geant4 (Red solid) and Penelope
(Green dot-dashed) Monte Carlo simulations.
\label{fig:nibf}}
\end{figure}

The data were integrated over angle to determine the total
normal-incidence backscattered fraction $\eta$.  The results of this
integration are shown in Fig.~\ref{fig:nibf} and are compared with
current integration measurements (described in
section~\ref{sec:current}), with previous data, and with the models.
In Fig.~\ref{fig:nibf}, the total systematic uncertainty, including
extrapolation to 50~eV and extrapolation over unmeasured angles, is
plotted.

Fig.~\ref{fig:nibf} compares the data with the Geant4 and Penelope
simulations.  The same discrepancies in normalization are again
observed.  Both Penelope and Geant4 adequately describe the reduction
of $\eta$ as the beam energy increases.

\subsection{Current Integration Mode\label{sec:current}}

The results for $\eta$ based on our current integration measurements
are also shown in Fig.~\ref{fig:nibf}.

The silicon detector measurements are found to be systematically
higher than the current mode measurements; however, the two methods of
are found to agree within the systematic uncertainties.  In the case
of the current integration method, this systematic uncertainty is
dominated by residual correction for secondary electron collection due
to the penetration through the grid of the target rod and
reproducibility of the experimental results under varying conditions
(see Table~\ref{tab:currmode}).  In the case of the silicon detector
measurements, it is dominated by reproducibility, and by uncertainties
in alignment, solid angle effects, and extrapolation to 50~eV (see
Table~\ref{tab:simode}).

The data are also compared with previous data on Be and Si targets due
to Drescher et al.~\cite{bib:Drescher} and with data on Be due to
Neubert et al.~\cite{bib:Neubert1}.  Both groups used current
integration techniques to arrive at their results.  Neubert et
al.~\cite{bib:Neubert1} in particular used a second target apparatus
to study the effects of secondary electrons, as opposed to the grid
used in this work.  Only the subsets of their data that overlap the
region 43.5 to 124 keV are plotted.  The Drescher data on Be are
systematically higher than the Neubert data.  As noted previously, our
data tend to agree with the Neubert data on Be, as do the data of
Massoumi et al.~\cite{bib:massoumi1,bib:massoumi2} taken below 40 keV.

Our data on organic plastic scintillator to our knowledge are the
first in this energy range.  As expected, the results lie below the
previous measurements on carbon (not shown), and above our data and
the Neubert data on beryllium.

\section{Quantitative Comparison to Models\label{sec:fit}}

Monte Carlo to data fits including a single fit parameter, a global
normalization factor, were performed for four different observables
measured by our experiments (including those presented in
Ref.~\cite{bib:prcme}).  The observables considered for fitting were:
$\frac{1}{N_e}\frac{dN}{d\Omega dq}$ at 124 keV beam energy,
$\frac{d\eta}{d\Omega}$ at 124 keV beam energy, $\eta$ from silicon
detector mode, and $\eta$ from current integration mode.  The
technique of $\chi^2$ minimization was used to constrain the fit.  As
mentioned in our previous work \cite{bib:prcme}, point-to-point
statistical and systematic uncertainties were exceedingly small
relative to the overall normalization systematic.  To simplify the
fitting procedure, it was assumed that the point-to-point
uncertainties were proportional to the global normalization
uncertainty, for the purposes of evaluating $\chi^2$.

The results of the fit are listed in Table~\ref{tab:final}.
\begin{table}
\begin{ruledtabular}
\begin{tabular}{ccccc}
Target & Observable & Geant4       & Penelope     & $\chi^2$ Ratio\\
       &            & Factor       & Factor       & (G4/Pen)\\\hline
Be     & $\frac{1}{N_e}\frac{dN}{d\Omega dq}$(124 keV)
                    & 0.74         & 1.10          & 3.7\\
       & $\frac{d\eta}{d\Omega}$(124 keV)
                    & 0.80         & 1.09          & 2.8\\
       & $\eta$(Si det.)
                    & 0.84         & 1.02          & 3.1\\
       & $\eta$(current int.)
                    & 0.75         & 0.91         & 2.2\\
\hline
Si     & $\frac{1}{N_e}\frac{dN}{d\Omega dq}$(124 keV)
                    & 1.00        & 1.08           & 1.7\\
       & $\frac{d\eta}{d\Omega}$(124 keV)
                    & 0.98        & 1.10           & 1.4\\
       & $\eta$(Si det.)
                    & 0.98        & 1.08           & 1.3\\
       & $\eta$(current int.)
                    & 0.83        & 0.91          & 1.3\\
\hline
Scint. & $\frac{1}{N_e}\frac{dN}{d\Omega dq}$(124 keV)
                    & 1.07         & 1.16           & 1.8\\
       & $\frac{d\eta}{d\Omega}$(124 keV)
                    & 0.94        & 1.23           & 4.8\\
       & $\eta$(Si det.)
                    & 0.92        & 1.12           & 0.74\\
       & $\eta$(current int.)
                    & 0.80        & 0.97          & 0.38\\
\end{tabular}
\end{ruledtabular}
\caption{Overall scale factors and $\chi^2$ ratios, comparing Geant4
to Penelope, under assumption of point-to-point uncertainty
proportional to estimated normalization systematic
uncertainty.\label{tab:final}}
\end{table}
The fit results for $\frac{1}{N_e}\frac{dN}{d\Omega dq}$(124 keV) data
for scintillator are displayed in Fig.~\ref{fig:120}.  Note that in
our previous work \cite{bib:prcme}, we did not use this fitting
technique, and normalization factors were simply determined by eye.
However, the normalization factors determined using the $\chi^2$
minimization method have a good correspondence with those numbers.

As the absolute point-to-point uncertainty was not determined
precisely, the values of uncertainties on the fit parameter, and of
absolute $\chi^2$'s, have no meaning.  Therefore only the ratio of
$\chi^2$'s determined for the Geant4 model, divided by that for the
Penelope model is quoted.  I.e., when taking the point-to-point
uncertainty to be equivalent to the total normalization uncertainty,
the value of the reduced $\chi^2$ was generally significantly less
than unity.  The exception to this was the comparison of the Geant4
simulation to $\frac{1}{N_e}\frac{dN}{d\Omega dq}$ for Be targets,
where an absolute reduced $\chi^2$ of 2.1 was seen for 188 degrees of
freedom.

Overall normalization scale factors determined for both Penelope and
Geant4 were generally found to agree within the normalization
systematic uncertainties with unity, although deviations up to 25\%
are seen in certain cases.  However, we note that, for observables
which are not susceptible to extrapolation uncertainties, namely
$\frac{1}{N_e}\frac{dN}{d\Omega dq}$(124 keV) and $\eta$(current
int.), the Penelope scale factors are globally within 16\% of unity.

Penelope generally gives a lower $\chi^2$ than Geant4, indicating that
the shape of the data is better described by Penelope.  The exception
is in the beam-energy dependence of $\eta$ for the scintillator data,
for which Geant4 gives a somewhat better fit.  However, we believe
this to be a coincidental cancellation, given that Geant4 gives a
worse description of the data for the other two observables for
scintillator targets.

The reason for the superior description of the data by Penelope is
likely due to its treatment of multiple scattering.  In the case of
backscattering, the multiple scattering effects are dominated by
large-angle scattering.  The cross-section for large-angle scattering
is dominated by Mott scattering.  Multiple scattering algorithms,
however, do not necessarily include all collisions of a given
particle.  Such algorithms are referred to as ``condensed''
algorithms, where algorithms which include all collisions are referred
to as ``detailed''.  Most particle physics simulation codes, such as
Geant4, use multiple scattering theories which are improved versions
of Moli\`ere theory, and are therefore condensed algorithms requiring
e.g. step sizes to be chosen very carefully~\cite{bib:geant4msc}.
Such algorithms generally perform adequately for small angle
scattering.  More recently, newer multiple scattering algorithms have
become available, known as ``mixed'' algorithms which simulate hard
collisions one by one (such as large-angle Mott scattering) and use a
condensed algorithm to calculate the effects of soft collisions (such
as small-angle Mott scattering and electron-electron scattering).
Penelope belongs to the mixed class of simulation codes.

We note that aspects of Penelope are included in the most recent
versions of Geant4.  However, the crucial aspect of Penelope for the
correct description of backscattering, which is the mixed algorithm
multiple scattering code, is to date not included in Geant4.

\section{Conclusion}

Our new data on scintillator answer important questions regarding the
systematic uncertainties due to backscattering for a broad range of
low-energy beta spectroscopy experiments.  Most importantly, the data
agree well with models implemented in the Geant4 and Penelope Monte
Carlo codes.

Overall normalization scale factors were determined using a $\chi^2$
minimiziation technique.  The resultant scale factors are generally
found to agree within the normalization systematic uncertainties with
unity.  In some cases, discrepancies of up to 25\% are seen.  However,
for observables which are not susceptible to extrapolation
uncertainties, the Penelope scale factors are always within 16\% of
unity.  In general, Penelope also gives lower $\chi^2$ values than
Geant4, indicating a better fit to the shape of the data in terms of
energy and angle of the backscattered electrons, and in terms of beam
energy.  The reason for the superior description of backscattering by
Penelope is likely due to its more accurate treatment of multiple
scattering, employing a mixed algorithm treating large-angle
scattering exactly, while using a condensed algorithm for small-angle
scattering.

\section{Acknowledgements}

We gratefully acknowledge the technical support of Robert Carr at the
California Institute of Technology.  This work was supported by the
National Science Foundation.  One of us (JWM) is supported by the
Natural Sciences and Engineering Research Council of Canada.

\bibliography{bs_paper}

\end{document}